\documentclass[manuscript]{aastex}

\usepackage{amsmath}
\usepackage{amssymb}
\usepackage{graphics}

\def\teff{$T_\mathrm{eff}$}
\def\logt{$\log T_\mathrm{eff}$}
\def\logg{$\log g$}
\def\logw{$\log W$}
\def\loggM{$\log g_\mathrm{model}$}
\def\loggF{$\log g_\mathrm{fit}$}
\def\Mv{${M_V}$}

\newcommand{\FeI}{\ion{Fe}{1}}
\newcommand{\FeII}{\ion{Fe}{2}}

\shorttitle{Wilson-Bappu Effect: Extended to Surface Gravity}
\shortauthors{Park et al.}

\begin{document}

\title{Wilson-Bappu Effect: Extended to Surface Gravity}

\author{Sunkyung Park$^1$, Wonseok Kang$^1$, Jeong-Eun Lee\footnote{\it{Corresponding Author :} Jeong-Eun Lee (jeongeun.lee@khu.ac.kr)}, Sang-Gak Lee$^2$} 
\affil{
       $^1$ School of Space Research, Kyung Hee University, Yongin-Si, Gyeonggi-Do 446-701, Republic of Korea; 
	    sunkyung@khu.ac.kr, wskang@khu.ac.kr, jeongeun.lee@khu.ac.kr  \\
       $^2$ Astronomy Program, Department of Physics and Astronomy, Seoul National University, Seoul 151-742, Republic of Korea;
	    sanggak@snu.ac.kr
}

\begin{abstract}

Wilson and Bappu found a tight correlation between the stellar absolute visual magnitude (\Mv) and the width of the Ca II K emission line for late-type stars in 1957.
Here, we revisit the Wilson-Bappu relationship (hereafter, WBR) to claim that WBR can be an excellent indicator of stellar surface gravity of late-type stars as well as a distance indicator. 
We have measured the width ($W$) of the Ca II K emission line in high resolution spectra of 125 late-type stars, which were obtained with Bohyunsan Optical Echelle Spectrograph (BOES) and adopted from the UVES archive.
Based on our measurement of the emission line width ($W$), we have obtained a WBR of \Mv\ = 33.76 - 18.08 \logw. 
In order to extend the WBR to be a surface gravity indicator, the stellar atmospheric parameters such as effective temperature (\teff), surface gravity (\logg), metallicity ([Fe/H]), and micro-turbulence ($\xi_{tur}$) have been derived from the self-consistent detailed analysis using the Kurucz stellar atmospheric model and the abundance analysis code, MOOG. 
Using these stellar parameters and \logw, we found that \logg\ = $-$ 5.85 \logw\ $+$ 9.97 \logt\ $-$ 23.48 for late-type stars. 

\end{abstract}

\keywords{Stars: late-type --- Stars: fundamental parameters --- Techniques: spectroscopic}

\section{Introduction}

The Ca II K line ($\lambda =$ 3933.7 \AA) of late-type stars has 
chromospheric emission feature superimposed on the deep photospheric absorption.
The Ca II K emission feature is stronger in more evolved stars since the chromospheric activity increases as a late-type star evolves from the main sequence to the giant stages (i.e., as the luminosity increases) \citep{pasquini90, dupree99}.

\citet{wilson57} found a strong relationship between the absolute visual magnitude (\Mv) and the width of the Ca II K emission line (\logw) of late-type stars, referred to as the Wilson-Bappu relation (WBR); one feature of this relation is that it is independent of stellar spectral type.
\citet{wallerstein99} obtained the WBR using $Hipparcos$ parallaxes for the first time.
\citet{pace03} used high resolution spectra and $Hipparcos$ data to obtain a WBR, and they
applied the WBR to estimate the distance to M67.

Several studies regarding the relationship between the width of the Ca II K emission line and the stellar parameters, 
such as effective temperature (\teff), surface gravity (\logg), and metallicity ([Fe/H]) have been carried out over the past four decades \citep{reimers73, neckel74, ayres79, lutz82}.
\citet{reimers73} found empirically that the width of the Ca II K emission line is related to \teff\ and \logg\, and \citet{neckel74} provided a theoretical framework for this relation.
\citet{ayres79} showed how the width of the Ca II K emission line varies with the surface gravity ($\Delta \lambda_{HWHM} \sim g^{-1.4}$) and \teff ;
the width is sensitive to the gravity, but insensitive to \teff.
\citet{lutz82} also showed that the width of the Ca II K emission line depends on \teff, \logg, and [Fe/H].
\citet{dupree95} also reported that the width of the Ca II K emission line depends on [Fe/H] in metal-poor stars with [Fe/H] = -2 $\sim$ -3.
However, \citet{gomez12} re-examined the dependence of [Fe/H] on WBR and they concluded that WBR is insensitive to metallicity.

Recently, late-type stars, especially M dwarfs, have been a subject of interest
because planets in the habitable zone can be more easily detected in late-type (low-mass) stars. 
The chemical composition of host stars is an important connection between the properties of planets and the environment in which they formed.
Conventionally, the chemical abundances of late-type stars are estimated using an abundance analysis code and a model stellar atmosphere, which requires the information of \teff, \logg, and [Fe/H].
However, stellar atmospheric models require spectra covering a sufficient number of Fe I and Fe II lines.
In addition, singly ionized atomic absorption lines such as \FeII\ are not easily detected in M type stars, including M dwarfs.
\teff\ of late-type stars can be derived relatively easily using spectroscopic or photometric methods 
\citep[e.g.][]{belle99, casagrande10}.
However, a simple indicator of stellar gravity has not yet been proposed.

In this context, we extend WBR to be an indicator of the surface gravities of late-type stars, including M type stars.
This is done for the first time, although there have been many studies regarding WBR over the half-century.

In Section 2, we describe our observations and spectral data.
In Section 3, we explain our method for determining the parameters of stellar atmospheres and measuring the width of the observed Ca II K emission lines. 
We present our results and discuss WBR as a surface gravity indicator, in Sections 4 and 5, respectively.
We summarize our results in Section 6.

\section{The Data}

\subsection{BOES Observations}

The Bohyunsan Optical Echelle Spectrograph (BOES) is a high resolution echelle spectrograph 
attached to a 1.8-m optical telescope at Bohyunsan Optical Astronomy Observatory (BOAO).
41 G and K type stars were observed using BOES in 2008 and 2009 \citep{kang11} and 31 G, K and M type stars were observed during April 2-5 and May 21-23 in 2012 at spectral resolution ($  R = \frac{\lambda} {\Delta\lambda} $) of 45,000 using the 200 $\mu$m fiber. 
The observed wavelength range was  3600\AA{} -- 10500\AA{}, covering the full optical spectrum. 
The signal-to-noise ratio (SNR) at the bottom of the Ca II K emission line is typically about 30, ranging from 12 to 107. 
The information on the observed targets are listed in Table~\ref{tbl_total}.

The observation data were reduced by the IRAF (Image Reduction and Analysis Facility) {\tt echelle} package.
Each aperture from the spectral images was extracted using a master flatfield image. 
Using the flatfielding process, we corrected the interference fringes and pixel-to-pixel variations of the spectrum images.
A ThAr lamp spectrum was used for wavelength calibration.

\subsection{UVES Archive} 

Echelle spectra of 53 late-type stars (G, K and M) were adopted from the UVES POP (Paranal Observatory Project) field star archive \citep{bagnulo03}. 
UVES POP provides data in a reduced form.
The spectra were obtained using the UVES echelle spectrograph at the Very Large Telescope (VLT) by the European Southern Observatory (ESO). 
To cover the full optical wavelength range (304 -- 1040 nm), all stars were observed with two instrument modes: Dichroic $\#$1 and Dichroic $\#$2.
We used the Dichroic $\#$2 437 blue arm data, which has a central wavelength of 437 nm and covers 373 to 499 nm, where the Ca II K emission line is located.

UVES provides high resolution and efficiency at UV wavelengths.
The UVES data have a spectral resolution of $\sim$ 80,000 and median SNR of $\sim$ 100 at the bottom of the Ca II K emission line; the SNR of the UVES spectra ranges from 32 -- 246.
The stars included in this work are listed in Table~\ref{tbl_total}.

\section{Spectroscopic Analysis}

\subsection{Stellar Atmospheric Parameters}

By using a model atmosphere and abundance analysis code, we determined \teff, \logg, [Fe/H] and micro-turbulence ($\xi_{tur}$) for the stellar atmosphere.
We performed a self-consistent detailed analysis using the equivalent widths ($EW$s) of the \FeI\ and \FeII\ lines.
The $EW$s of the \FeI\ and \FeII\ lines were measured using TAME (Tool for Automatic Measurement of Equivalent-width) \citep{kang12} for the 80 G and K type stars, whose stellar properties are appropriate to use the adopted stellar atmospheric model (e.g. \teff $\ge$ 3500 K).
From the $EW$s data, the Fe abundance from each \FeI\ and \FeII\ line was estimated by the revised 2010 version of MOOG \citep{sneden73} and Kurucz ATLAS9 model grids \citep{kurucz93}.

We chose the effective temperature and the micro-turbulence that minimized the slope of [Fe/H] derived from the individual lines as a function of, respectively, 
the excitation potentials (EP) and the $EW$s of the lines.  
Then we chose the surface gravity that minimized the difference between iron abundances derived from \FeI\ and \FeII\ lines 
(\textquotedblleft$\log g_{model}$\textquotedblright is used to refer to the gravity derived by the stellar atmospheric model hereafter).
We selected the average of the Fe abundances derived from the \FeI\ and \FeII\ analysis as the metallicity.
The method is iterative (find more details in Kang et al. 2011).
We iterated until the slopes of \FeI\ abundance vs. EP and $EW$ become smaller than $10^{-5}$, and the difference between the abundances from \FeI\ lines and from \FeII\ lines becomes smaller than 0.003 dex.
Through this procedure, we determined the final stellar atmospheric parameters of the 80 G and K type stars, and their stellar atmospheric parameters are listed in Table~\ref{tbl_total}.

\subsection{Measurement of the Emission Width of a Ca II K Line}

In the past, both Ca II H and K lines were used to derive WBR because of the low spectral resolution.
However, in high resolution spectroscopy, it has been revealed that the Ca II H line is contaminated by other adjacent lines, and thus only the Ca II K line can be used for WBR \citep{wilson76}. 
Hence, we also used Ca II K emission line widths to derive the WBR of our samples.

\begin{figure}[!tb]
 \plotone{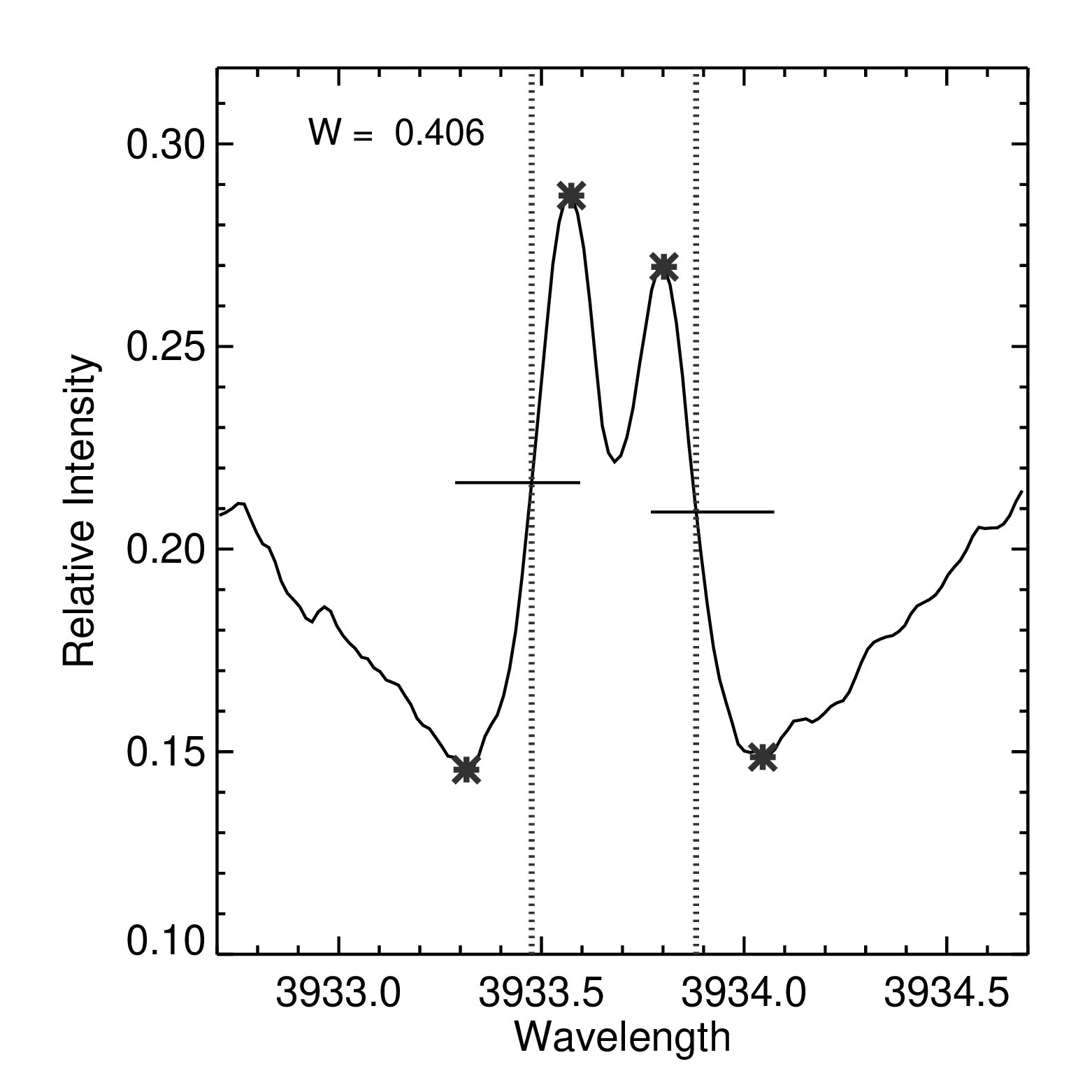}
 \caption{Measurement of Ca II K emission line widths. 
	  Asterisks mark the maximum and the minimum points of red and blue peaks.
	  The difference between the two dashed lines represents the width of the Ca II K emission line (\logw).}
 \label{idl_measure} 
\end{figure}

We measured the widths of the Ca II K emission lines for our samples using the method from \citet{pasquini92}, \citet{dupree95}, and \citet{pace03}.
The width of the Ca II K emission line is defined as the wavelength difference between the half intensities of two emission peaks (dashed lines in Fig.~\ref{idl_measure}). 
The widths calculated by this method show better correlation between \Mv\ and \logw\ than those of Wilson's method, which uses the wavelength difference between two minimum points.
We define the errors of the line widths (1) as the standard deviation for the stars observed more than three times or (2) as the half difference between the largest and the smallest values measured for the stars with 2 or 3 spectra.
The UVES data are observed at least twice. 
However, only three stars were observed more than once with BOES.
Therefore, for the three stars, we applied the same method as used for the UVES data.
Then, the average value of the three errors was adopted for the stars observed with BOES only once.
Measured widths of the Ca II K emission lines are listed in Table~\ref{tbl_total}.

\citet{lutz70} defined $\rm{W_{0}}$ as the width at half-maximum of the emission profile, with 
a correction applied to account for the resolution limit of the spectrograph itself, which may  broaden the lines.
\citet{dupree95} and \citet{pace03} did not correct for instrumental broadening because this effect was much smaller than the observed line width of the emission component.
The predicted width of the BOES instrumental profile is about 0.1 \AA{},
which we estimated from the emission lines in the comparison lamp spectrum and is 5 $\sim$ 28 $\%$ of the line widths of the Ca II K emission line of our samples (the median value is $\sim$ 15 $\%$).
The width of the UVES instrumental profile is about 0.15 $\sim$ 0.20 \AA{} \citep{cox05}, but the data are provided as a reduced form, which are almost free from the instrumental profile.
As a result, we ignore the instrumental broadening effect following \citet{dupree95}.

\section{Results}

\subsection{The Wilson-Bappu Relation}

The absolute visual magnitudes (\Mv) were calculated from apparent visual magnitudes ($m_{V}$) and trigonometric parallaxes, both of which are taken from the ${Hipparcos}$ Catalogue \citep{leeuwen07}.
In order to minimize the absolute magnitude error, the samples were limited to stars which have parallax errors less than 10$\%$. 
As a result, 125 stars were selected, and are listed in Table \ref{tbl_total}, along with their
calculated absolute magnitudes.

\begin{figure}[!tb]
 \plotone{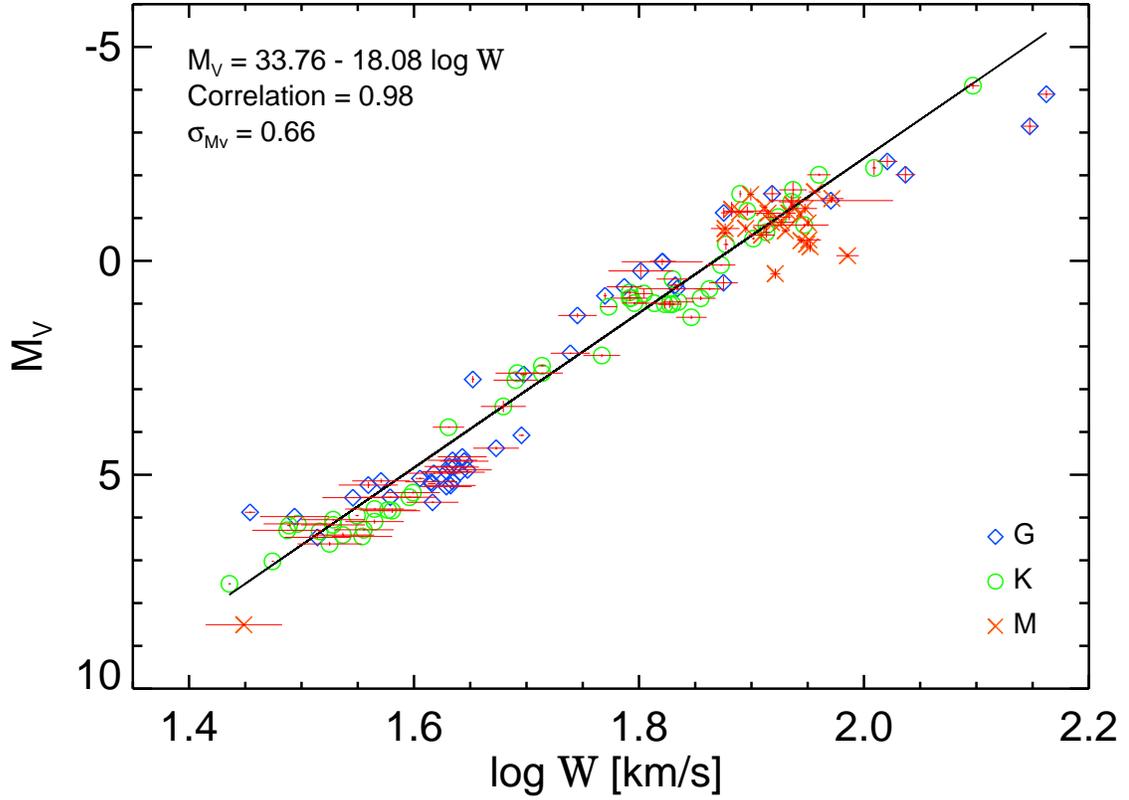}
 \caption{The Wilson-Bappu relation of our samples, \Mv\ vs. \logw. The 125 stars with parallax errors less then 10$\%$ were used for fitting.
  The blue diamond, green circle, and orange cross symbols indicate G, K, and M type stars, respectively. 
  The red error bars represent the errors of the measurements of \logw\ and \Mv\ that originated from their parallax measurements.} 
   \label{Mv_logW}  
\end{figure}

Using \Mv\ and the width of the Ca II K emission line (\logw) (Table~\ref{tbl_total}), we obtained the WBR by a linear least squares fit weighting of the \Mv\ errors.
In this fitting, we considered the Lutz-Kelker effect;
\citet{lutz73} noted that there is a systematic error in absolute magnitude as the observed parallaxes are larger than the true parallaxes on average.
This systematic error is related to the ratio of parallax error to parallax ($\frac{{\sigma}_{\pi}}{\pi}$) (see Table 1 in \citet{lutz73} or Table~\ref{tbl_LKE} in this paper).
104 stars in our samples have $\frac{{\sigma}_{\pi}}{\pi}$ less than 0.050 indicating that the correction for $M_{V}$ is $\sim$ 0.02 mag.
Only 8 stars in our samples have $\frac{{\sigma}_{\pi}}{\pi}$ larger than 0.075, for which 0.11 mag must be applied to correct for $M_{V}$.
As a result, our WBR is
\begin{equation} \label{eq_WBR_LKE}
M_{V} = 33.76 - 18.08 \log W,
\end{equation}
as shown in Fig.~\ref{Mv_logW}.
\Mv\ and \logw\ show a very tight correlation, with a Pearson's correlation coefficient of 0.98, and the standard deviation is 0.66.

Extinction can affect the absolute magnitudes of stars.
However, the distance to most of our sources (114 stars) is less than 200 pc.
When we used only the 114 stars within 200 pc to minimize the extinction effect as in \citet{pace03}, the derived WBR 
was not very different (0.03 mag) from our original WBR (Eq.~(\ref{eq_WBR_LKE})).

\begin{figure}[!p]
 \plotone{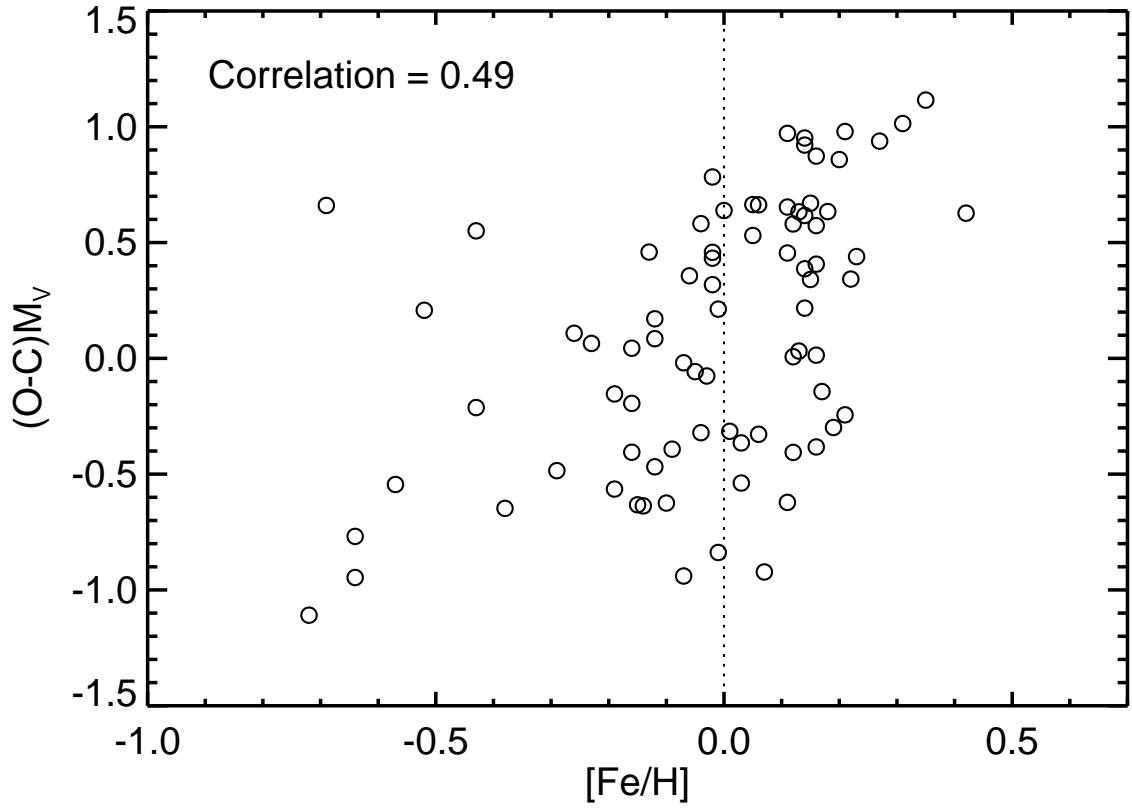}
 \caption
  {(O-C)Mv vs. [Fe/H]. 
  The correlation coefficient between two quantities is 0.49. Dashed line represents the solar metallicity.}
 \label{check_abu_allstars}
\end{figure}

We also examined the effect of metallicity in our WBR.  The metallicity of our samples covers a wide range (-0.72 $\le$ [Fe/H] $\le$ 0.42), although the majority of our sources are close to solar metallicity (Fig.~\ref{check_abu_allstars}).
The correlation coefficient between (O-C)Mv, which is the difference in Mv between the observed data point and the linear fit, and [Fe/H] is 0.49 for all of our sources, and it is 0.29 for the stars that have metallicity smaller than 0.0.
Therefore, we conclude that there is no metallicity dependence of the WBR in our sample, as noted in previous studies \citep{wilson57, gomez12}.

In order to evaluate our measurements, we compared our $W$s with those in previous studies by \citet{pace03} and \citet{wallerstein99}. 
The data used in \citet{pace03} were obtained under very similar conditions to our sample; 
they used high resolution spectra (R $\sim$ 60,000), parallaxes and visual magnitudes from the ${Hipparcos}$ Catalogue (ESA 1997), and same definition of the width of the Ca II K emission line.
However, the data from \citet{wallerstein99} were obtained under different conditions from ours;
they used $Hipparcos$ parallaxes (ESA 1997) and Ca II K emission line width adopted from \citet{wilson76}, 
where $W_{0} = W - 18$ [km/s], taking into account the instrumental width.

\begin{figure}[!p]
 \plotone{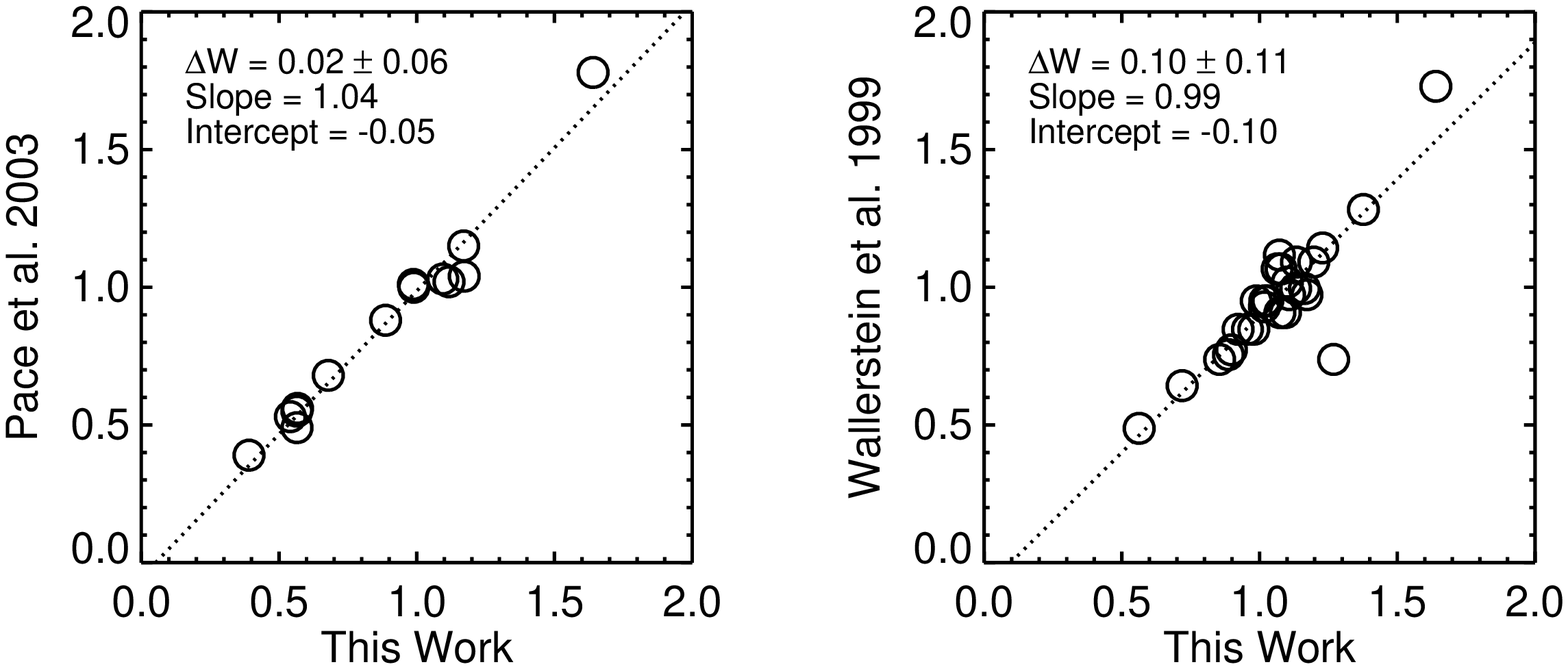}
 \caption{Comparisons with previous works for the width ($W$ [\AA]) of the Ca II K emission line: \citet{pace03} (left) and \citet{wallerstein99} (right). }
 \label{comp_prev}
\end{figure}

\begin{figure}[!p]
 \plotone{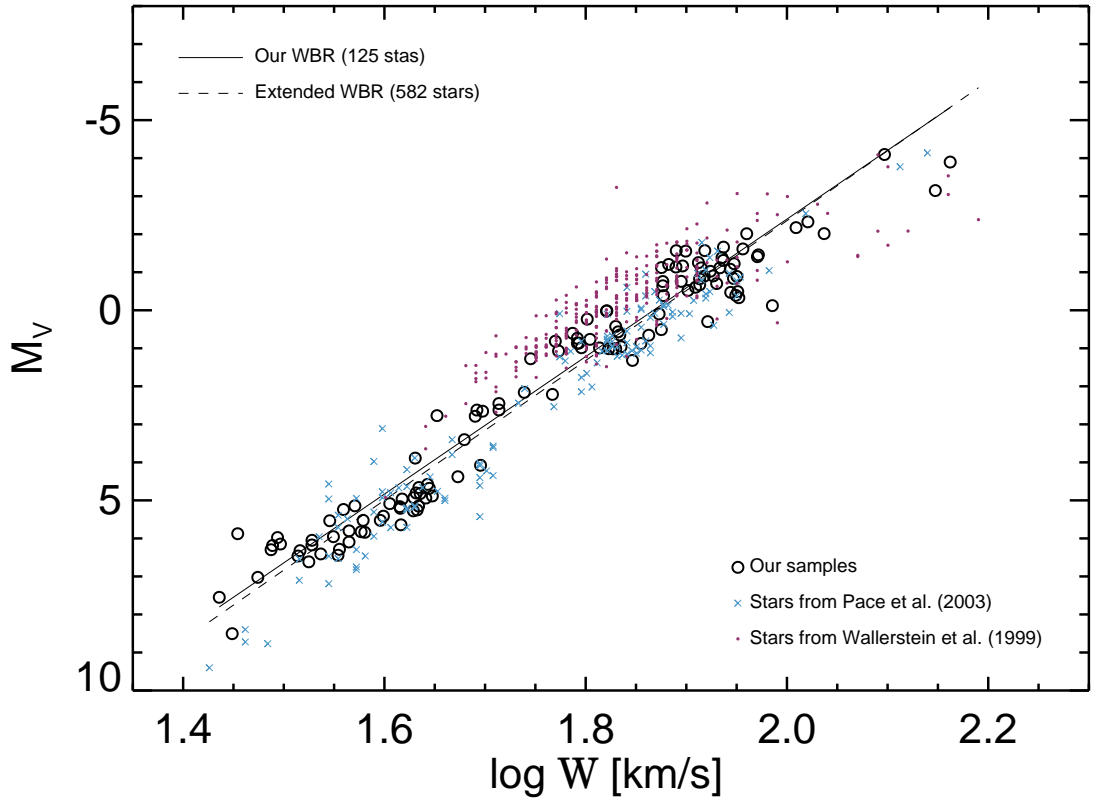}
 \caption{WBRs from extended and homogenised samples. Open circles indicate our sample (125 stars), blue crosses represent the stars from \citet{pace03} (131 stars), 
and purple filled circles coresspond to the stars from \citet{wallerstein99} (326 stars). 
  Solid line and dashed line represent our WBR of 125 stars ($M_{V}$ = 33.76 - 18.08 $\log W$) and the WBR ($M_{V}$ = 34.41 - 18.38 $\log W$) calculated from the combined sample (582 stars), respectively.}
 \label{comp_WBR}
\end{figure}

We have 14 and 28 stars in common with \citet{pace03} and \citet{wallerstein99}, respectively.
The mean differences of $W$ between our measurements and their values are $\sim0.02$ \AA\ and 0.10 \AA, and the standard deviations are $\sim0.06$ \AA\ and 0.11 \AA, respectively, as seen in Fig.~\ref{comp_prev}.
The relations between our widths and those measured by \citet{pace03} or \citet{wallerstein99} have slopes equal to 1 within the errors. 
Therefore, we corrected widths from \citet{pace03} and \citet{wallerstein99} by subtracting the intercept value of -0.05 \AA\ and -0.10 \AA\ found in Fig.~\ref{comp_prev} to make a homogeneous data sample when combined with ours.
We included 131 stars from \citet{pace03} and 326 stars from \citet{wallerstein99}, which are not in our list and have parallax errors smaller than 10$\%$.
We calculated \Mv\ for these 457 stars using ${Hipparcos}$ Catalogue \citep{leeuwen07} for data homogeneity.
As a result, we obtained the WBR from the 582 stars as $M_{V} = 34.41 - 18.38 \log W$ (dashed line in Fig.~\ref{comp_WBR}).
The median difference is about 0.20 between \Mv\ calculated by our WBR (125 stars) and the WBR with the extended sample (582 stars).
Therefore, this test supports that our WBR is well-calibrated.

\subsection{Emission Width as a Surface Gravity Indicator}

\logw\ has a tight relationship (WBR) with \Mv, which is associated with effective temperature and stellar radius;
\Mv\ $\propto \log L \propto \log (R_{*}^{2} T_{eff}^{4}) \propto \log (M_{*} g^{-1} T_{eff}^{4}) = \log M_{*} -\log g + 4\log T_{eff}$, where $M_{*}$ is the stellar mass. 
As a result, because $L \propto M_{*}^{\gamma}$, $M_{V} \propto \log W \propto \log L \propto \alpha\log g + \beta\log T_{eff}$, where $W$ is the width of the Ca II K emission line.
Hence, we revisit WBR in order to derive the surface gravities of late-type stars based on homogeneous high spectral resolution observations, using a consistent analysis across the sample.

\begin{figure}[!tb]
 \plotone{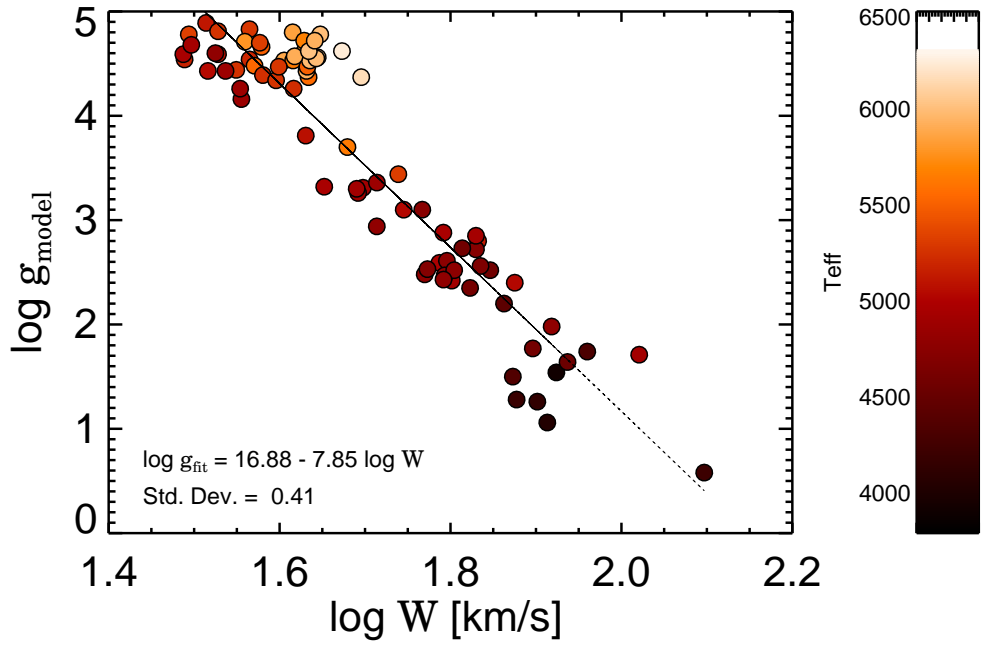}
 \caption{Plot of \logw\ vs. \loggM. 
	      Color represents the effective temperature as determined by detailed analysis with model atmospheres.
          The standard deviation of \logg\ is 0.41.}
 \label{logW_logg_teff}
\end{figure}

Using 80 G and K type stars with determined atmospheric parameters, first, we derived the relation between the width of the Ca II K emission line (\logw) and surface gravity (\loggM) as \loggF\ = 16.88 - 7.85 \logw\ (Fig.~\ref{logW_logg_teff}). 
However, as expected, \logw\ varies with temperature at a given gravity, as seen for \loggM\ $>$ 4.0 in Fig.~\ref{logW_logg_teff} (a smaller \logw\ at a lower temperature).
Therefore, we take into account \teff\ in order to determine the relationship between \loggM\ and \logw.

\begin{figure}[!tb]
 \plotone{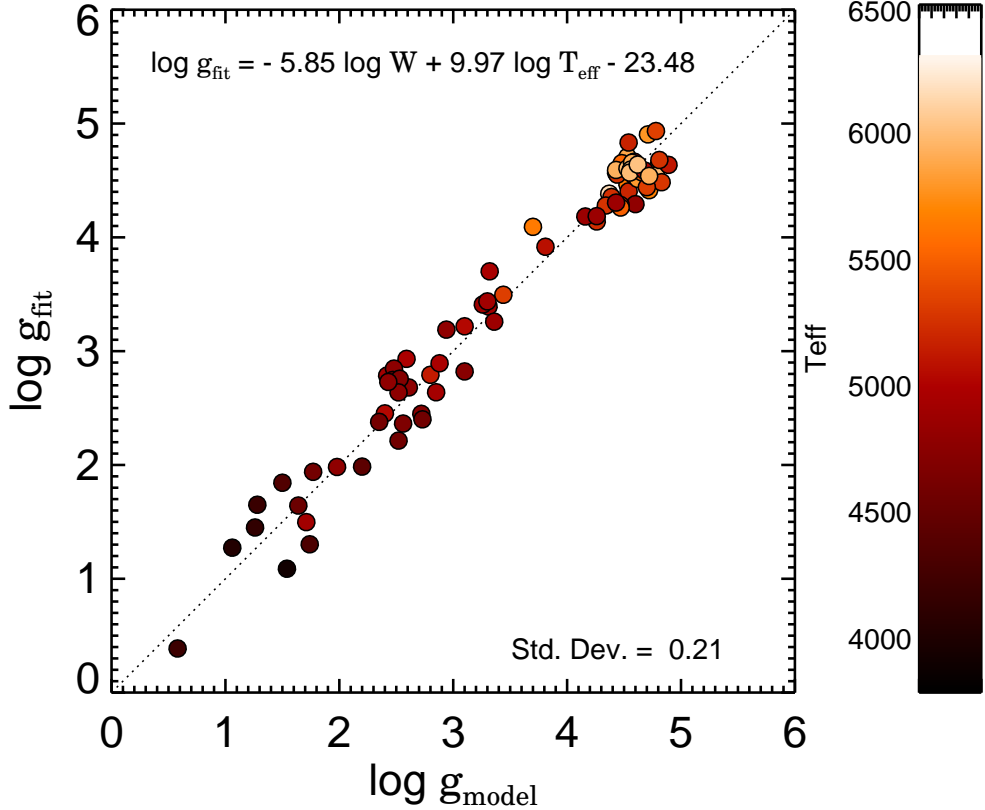}
 \caption{Plot of $\log g_\mathrm{model}$ vs. $\log g_\mathrm{fit}$.
	  $\log g_\mathrm{model}$ was derived from the stellar atmospheric model and
	  $\log g_\mathrm{fit}$ was obtained from Eq.~(\ref{eq_final}). 
	  Symbol color represents the effective temperature of the model atmosphere. 
	  The standard deviation in \logg\ is 0.21, which is smaller than that in Fig.~\ref{logW_logg_teff}.}
 \label{regr_comp}
\end{figure}

From a linear regression analysis of \loggM\ with \logw\ and \logt, we found that \loggF\ has the following strong relationship with \logw\ and \logt,
  \begin{equation}   \label{eq_final}
    \log g_\mathrm{fit} = - 5.85 \log W + 9.97 \log T_\mathrm{eff} - 23.48.
  \end{equation}
Fig.~\ref{regr_comp} shows the relation between the model-determined surface gravity, $\log g_\mathrm{model}$, and the surface gravity estimated by Eq.~(\ref{eq_final}), $\log g_\mathrm{fit}$.
The standard deviation of the differences between \loggM\ and \loggF\ is 0.21 dex.

Therefore, Eq.~(\ref{eq_final}) can be used directly in order to estimate the stellar surface gravity of a late-type star within this uncertainty
when the width of the Ca II K emission line is measured, and the effective temperature of the star is known.

\section{Discussion}

\subsection{Application of WBR to the Distance}

\begin{figure}[!p]
 \plotone{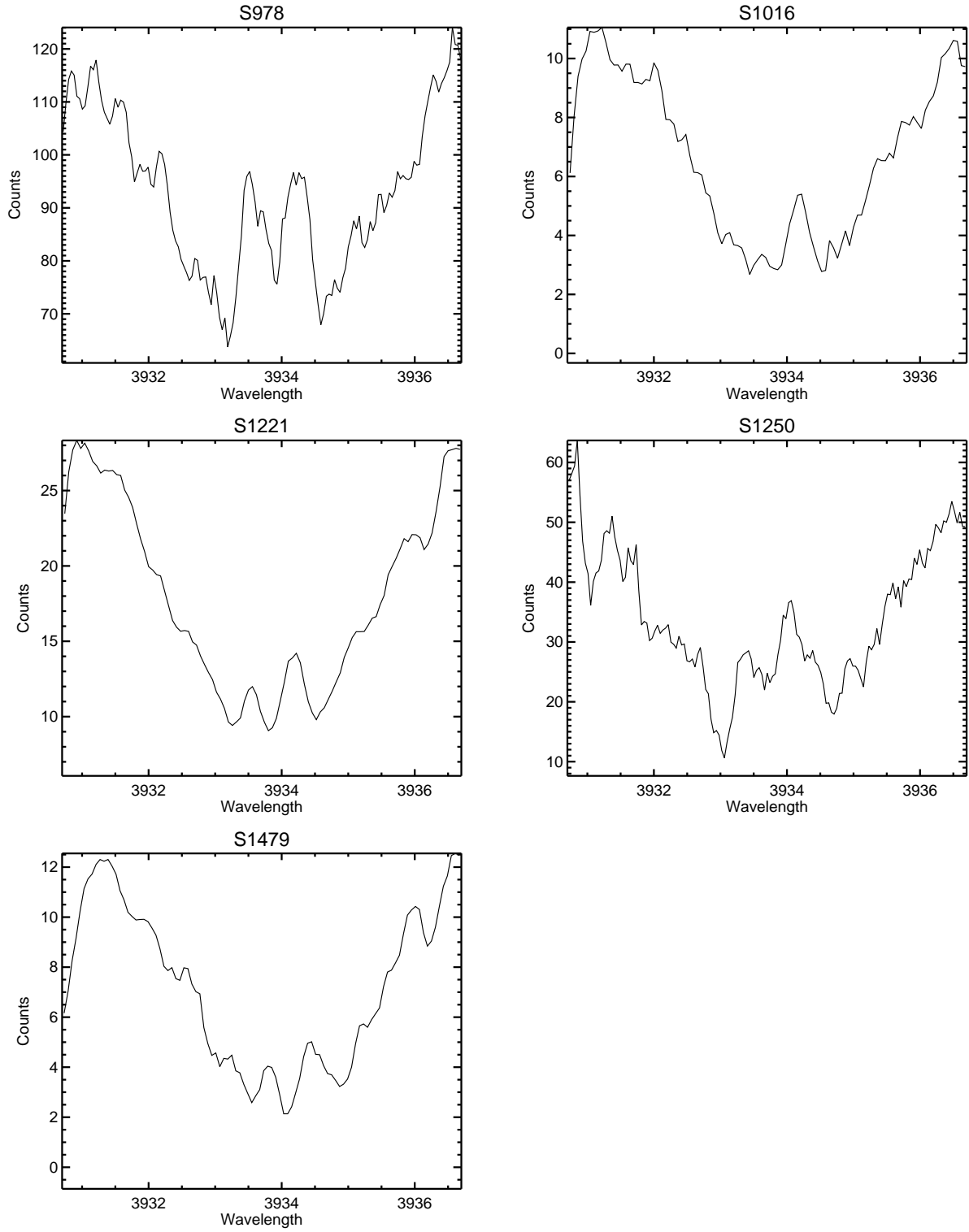}
  \caption[M67 spectra.]
   {M67 spectra. Five spectra from \citet{dupree99} were used to determine the distance modulus to M67 (S978, S1016, S1221, S1250, S1479).}
 \label{comp_M67}
 \end{figure}

WBR can be used as a distance indicator since the absolute visual magnitude can be calculated with the Ca II K emission line width. 
We applied our WBR to the open cluster M67 in order to calculate its distance modulus by using the spectra in \citet{dupree99}. 
\citet{dupree99} used Ca II H $\&$ K emission lines with intermediate resolution (R $\sim$ 30,000) echelle spectra of 15 red giants in M67.
We have adopted 5 of the best quality spectra (Sanders ID numbers : 978, 1016, 1221, 1250, 1479), shown in Fig.~\ref{comp_M67}. 
As noted in \citet{pace03}, we could not find the apparent Ca II K emission lines
from the following Sanders ID numbers : 258, 989, 1279, and 1316.
The comparison between our results and those of \citet{pace03} is listed in Table~\ref{tbl_M67}.

Based on their WBR, \citet{pace03} calculated the average distance modulus of M67 as 9.65 mag ($\pm$ 0.2 mag).
In previous studies \citep{montgomery93, carraro96, sarajedini09}, the range of the average of M67  is 9.55 $\leq$ $\rm{(m-M)_{V}}$ $\leq$ 9.85 mag.

To estimate the distance modulus to M67 using our WBR ($M_{V}$ = 33.76 - 18.08 $\log W$), we measured the widths of the Ca II K emission lines for the 5 stars listed above.
The absolute magnitudes of the stars calculated with our WBR are listed in Table~\ref{tbl_M67}.
The mean distance modulus of the 5 stars is 9.86, which is similar to 9.90, the mean distance modulus of the same 5 stars calculated by \citet{pace03}.
Our distance modulus also agrees well with values derived in previous studies \citep{montgomery93, carraro96, sarajedini09}.

\subsection{Comparison with \citet{lutz82}}

\begin{figure}[!tb]
 \plotone{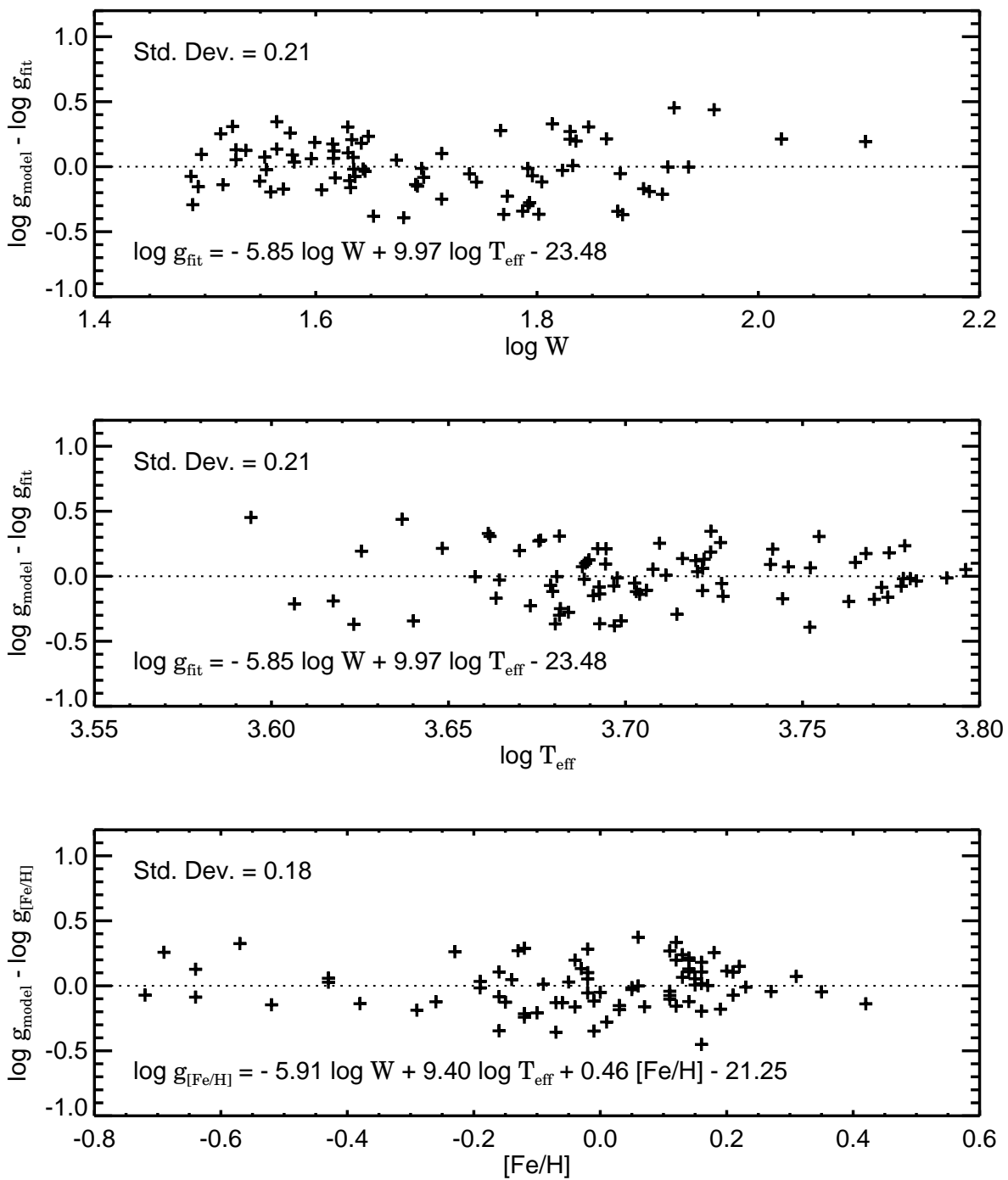}
 \caption{Surface gravity differences as a function of the width of the Ca II K emission line (\logw), effective temperature (\logt), and metallicity ([Fe/H]). The inclusion of [Fe/H] to \logg\ does not improve the fitting much.}
 \label{regr_comp_abu}
\end{figure}

\citet{lutz82} suggested that the width of the Ca II K emission line is related to stellar metallicity ([Fe/H]) as well as \logg\ and \teff.
Although the majority of our stars are comparable to the solar metallicity, our full sample spans a wide range of metallicity (-0.72 $\leq$ [Fe/H] $\leq$ 0.42).
If we fit \logw\ with all three stellar parameters (\logg, \teff, and [Fe/H]), the obtained relation is 
  \begin{equation}   \label{eq_abu}
    \log g_\mathrm{[Fe/H]} = - 5.91 \log W + 9.40 \log T_\mathrm{eff} + 0.46[Fe/H] - 21.25.
  \end{equation}
The metallicity term (0.46[Fe/H]) is significantly smaller than those of the width (5.91logW) and the temperature (9.40Teff), and it ranges from 
-0.33 to 0.19 in our sample, which is not much different from the standard deviation ($\sim$ 0.21) in the fitting of \logg\ (see Fig.~\ref{regr_comp_abu}).
As a result, we conclude that the dependence of [Fe/H] on \logg\ can be ignored.
(This is consistent with what we found in Section 4.1;  the role of metallicity is marginal in the WBR.)

The equation derived by \citet{lutz82} for stars with solar metallicity (-0.35 $\leq$ [Fe/H] $\leq$ 0.35) is  
  \begin{equation}  \label{eq_lutzW}
    \log W = -0.232 \log g + 1.78 \log T_\mathrm{eff} - 4.15, 
  \end{equation}
We can rearrange Eq.~(\ref{eq_lutzW}) to derive \logg\ as
  \begin{equation}   \label{eq_lutzG}
    \log g = - 4.31 \log W + 7.67 \log T_\mathrm{eff} - 17.9.
  \end{equation}

As noted above, the majority (86$\%$) of our sources are close to solar metallicity, and even including the full sample, the dependence of \logg\ on [Fe/H] is very minor.
Therefore, we can directly compare the Eq.~(\ref{eq_final}) with Eq.~(\ref{eq_lutzG}).
The mean difference in \logg\ between these two equations is $\sim$ 0.31, applied across 
the full sample.
\citet{lutz82} used low resolution spectral data and stellar atmospheric parameters from the literature; we have derived \logg\ of each sample star using a consistent detailed analysis with homogeneous  high resolution spectral data.
Therefore, we believe that the equation of gravity derived in this study is more robust.

\subsection{Application to M Type Stars} 

In this study, only G and K type stars were used to determine the relationship among \logg, \logt, and \logw\ because of temperature limits in the Kurucz ATLAS9 model grids (\teff $\ge$ 3500 K). 
In order to apply this relation to M type stars, the effective temperatures of M type stars need to be calculated by other methods. 
We applied the (V-K) color method to G and K stars, whose temperatures are derived using the stellar atmospheric model, and found that the method works well.
Following \citet{belle99}, we assume that the (V-K) color method is a reliable temperature indicator for M stars, as checked for G and K stars.
With the derived \teff\ and the measured width of the Ca II K emission line, \logg\ can be calculated for each M star based on the Eq.~(\ref{eq_final}).
The results are listed in Table~\ref{tbl_Mstar}. 
The calculated \teff\ and \logg\ are comparable to the values found in other studies \citep[][]{mallik98, smith86, koleva12}.
The difference between our calculations and the values reported in other studies is less than the standard deviation of \loggF\ ($\sigma_\mathrm{\log g, fit}$ = 0.21).
Therefore, we conclude that the relation determined in this study can be a useful tool to calculate the \logg\ of late-type stars without using stellar atmospheric models.

\section{Summary}

1. We derived the WBR ($M_{V}$ = 33.76 - 18.08 $\log W$) for 125 late-type stars, whose parallax errors are smaller than 10$\%$, using high resolution echelle spectra obtained from BOES and UVES and visual magnitudes and parallaxes taken from the $Hipparcos$ Catalogue.

2. Our WBR seems well-calibrated when compared to previous studies \citep{wallerstein99, pace03}.

3. We applied the WBR to M67 to calculate its distance modulus by using the spectra from \citet{dupree99}.
Our mean distance modulus from 5 stars agrees well with previous results.
Therefore, we believe that our WBR is well-calibrated. 

4. We determined the stellar atmospheric parameters (\teff, \logg, [Fe/H], and $\xi_{tur}$) with a stellar model atmosphere (Kurucz ATLAS9) and an abundance analysis code (2010 version of MOOG).  
Using \logg\ and \logt\ determined by the model, and the measurements of the Ca II K emission line width (\logw), we found a relation of $\log g_\mathrm{fit}$ = - 5.85 $\log W$ + 9.97 $\log T_\mathrm{eff}$ - 23.48 with the standard deviation of 0.21 dex.
 
5. The surface gravities calculated with the above equation for 4 M type stars agree well with those values derived in previous studies within the standard deviation of $\log g_{fit}$ ($\sigma_\mathrm{\log g, fit}$ = 0.21). 
Therefore, this relation can provide a simple way to calculate the surface gravity of late-type stars without using stellar atmosphere models.

\acknowledgements
We are grateful to Dr. Dupree who kindly provided the M67 data. We also thank the anonymous referee for constructive comments and Dr. Green for English editing and clarity.
This study was supported by the World Class University (WCU) program (No. R31-10016) and the Basic Science Research Program (No. 2012-0002330)
through the National Research Foundation of Korea (NRF) funded by the Ministry of Education, Science and Technology (MEST) of the Korean government.
This work was also supported by the Korea Astronomy and Space Science Institute (KASI) grant funded by the Korea government(MEST).

\clearpage

\clearpage


\begin{center}
\begin{deluxetable}{llccccccccccc} 
\footnotesize
\rotate
\tablecaption{Our data sample \label{tbl_total}}
\tablewidth{0pt}
\tablehead{
\colhead{HD name} & \colhead{Sp.type$^{b}$} & \colhead{V$^{c}$} &  \colhead{K$^{d}$} & \colhead{Width}   & \colhead{Werr}    & \colhead{$M_{V}$} & \colhead{$M_{V}$err} & \colhead{$T_{eff}$} & \colhead{$\log g$} & \colhead{[Fe/H]} & \colhead{$\xi_{tur}$}   & \colhead{Instrument} \\
\colhead{}        & \colhead{}              & \colhead{[mag]}   & \colhead{[mag]}    & \colhead{[\AA{}]} & \colhead{[\AA{}]} & \colhead{[mag]}   & \colhead{[mag]}      & \colhead{[K]}       & \colhead{[dex]}    & \colhead{[dex]}  & \colhead{[km s$^{-1}$]]} & \colhead{} 
}

\startdata

        HD3651$^{a}$ &             K0V &   6.03 &   4.00 &  0.482 &  0.029 &   5.80 &  0.008 &  5201 &  4.54 &  0.15 &  0.73 &       BOES \\
        HD7924$^{a}$ &              K0 &   7.31 &   5.16 &  0.443 &  0.029 &   6.17 &  0.017 &  5102 &  4.59 & -0.16 &  0.36 &       BOES \\
             HD11643 &            K1II &   6.26 &   3.55 &  0.873 &  0.010 &   1.01 &  0.104 &  4617 &  2.35 &  0.14 &  1.51 &       UVES \\
             HD11695 &           M4III &   4.41 &  -0.77 &  1.117 &  0.008 &  -0.70 &  0.044 &   -   &   -   &   -   &   -   &       UVES \\
             HD12642 &           M0III &   5.73 &   1.75 &  1.167 &  0.003 &  -0.39 &  0.105 &   -   &   -   &   -   &   -   &       UVES \\
       HD13974$^{a}$ &             G0V &   4.98 &   3.08 &  0.562 &  0.029 &   4.81 &  0.009 &  5944 &  4.43 & -0.43 &  0.33 &       BOES \\
       HD20367$^{a}$ &              G0 &   6.52 &   5.04 &  0.618 &  0.029 &   4.38 &  0.037 &  6253 &  4.62 &  0.16 &  1.25 &       BOES \\
       HD20630$^{a}$ &            G5Vv &   4.98 &   2.96 &  0.541 &  0.029 &   5.17 &  0.006 &  5860 &  4.80 &  0.14 &  1.00 &       BOES \\
             HD22049 &           K2Vk: &   3.87 &   1.78 &  0.431 &  0.001 &   6.32 &  0.002 &  5058 &  4.43 & -0.07 &  0.81 &       UVES \\
             HD23249 &        K1III-IV &   3.68 &   1.62 &  0.561 &  0.018 &   3.89 &  0.004 &  5080 &  3.81 &  0.16 &  0.88 &       UVES \\
             HD24616 &          G8IV/V &   6.83 &   4.55 &  0.589 &  0.001 &   2.77 &  0.079 &  4976 &  3.32 & -0.72 &  1.05 &       UVES \\
             HD25069 &             G9V &   6.00 &   3.60 &  0.654 &  0.004 &   2.66 &  0.041 &  4926 &  3.31 &  0.12 &  1.14 &       UVES \\
       HD25680$^{a}$ &             G5V &   6.04 &   4.38 &  0.583 &  0.029 &   4.89 &  0.012 &  6010 &  4.78 &  0.14 &  1.02 &       BOES \\
             HD27256 &        G8II-III &   3.50 &   1.44 &  0.869 &  0.023 &   0.01 &  0.011 &   -   &   -   &   -   &   -   &       UVES \\
       HD28305$^{a}$ &           K0III &   3.70 &   1.42 &  0.887 &  0.029 &   0.43 &  0.024 &  4949 &  2.85 &  0.21 &  1.42 &       BOES \\
       HD37124$^{a}$ &          G4IV-V &   7.79 &   5.95 &  0.488 &  0.029 &   5.14 &  0.051 &  5551 &  4.48 & -0.43 &  0.72 &       BOES \\
             HD37811 &           G7III &   5.60 &   3.36 &  0.868 &  0.072 &   0.02 &  0.076 &   -   &   -   &   -   &   -   &       UVES \\
       HD39587$^{a}$ &       G0VCH-0.3 &   4.52 &   3.00 &  0.567 &  0.029 &   4.82 &  0.005 &  5996 &  4.53 &  0.00 &  1.00 &       BOES \\
             HD42682 &        M2II-III &   5.59 &   0.72 &  1.041 &  0.007 &  -1.55 &  0.142 &   -   &   -   &   -   &   -   &       UVES \\
             HD45415 &           G9III &   5.71 &   3.20 &  0.772 &  0.003 &   0.81 &  0.068 &  4788 &  2.48 & -0.07 &  1.54 &       UVES \\
             HD48329 &            G8Ib &   3.19 &   0.12 &  1.906 &  0.029 &  -3.90 &  0.096 &   -   &   -   &   -   &   -   &       BOES \\
             HD59967 &             G3V &   6.79 &   5.10 &  0.529 &  0.005 &   5.09 &  0.018 &  5892 &  4.53 & -0.06 &  1.26 &       UVES \\
       HD65583$^{a}$ &             G8V &   7.11 &   5.09 &  0.409 &  0.029 &   5.98 &  0.020 &  5340 &  4.78 & -0.64 &  0.34 &       BOES \\
             HD67594 &            G2Ib &   4.52 &   2.32 &  1.843 &  0.029 &  -3.15 &  0.190 &   -   &   -   &   -   &   -   &       BOES \\
       HD72905$^{a}$ &          G1.5Vb &   5.76 &   4.17 &  0.544 &  0.029 &   4.96 &  0.012 &  5920 &  4.57 & -0.02 &  1.08 &       BOES \\
             HD74006 &         G7Ib-II &   4.13 &   1.84 &  1.226 &  0.156 &  -1.41 &  0.053 &   -   &   -   &   -   &   -   &       UVES \\
       HD75732$^{a}$ &             G8V &   6.11 &   4.01 &  0.543 &  0.029 &   5.64 &  0.020 &  5246 &  4.26 &  0.35 &  0.89 &       BOES \\
       HD76151$^{a}$ &             G3V &   6.14 &   4.46 &  0.559 &  0.029 &   4.93 &  0.015 &  5820 &  4.62 &  0.13 &  0.96 &       BOES \\
             HD76827 &           M3III &   4.81 &   0.31 &  1.269 &  0.029 &  -0.12 &  0.052 &   -   &   -   &   -   &   -   &       BOES \\
             HD77020 &            G9II &   6.05 &   3.43 &  1.087 &  0.020 &  -1.57 &  0.200 &  4793 &  1.98 & -0.14 &  2.06 &       UVES \\
             HD77912 &          G8Iab: &   4.73 &   2.40 &  1.377 &  0.029 &  -2.33 &  0.136 &  4922 &  1.71 & -0.13 &  2.26 &       BOES \\
             HD79354 &           K5III &   5.40 &   1.42 &  1.131 &  0.029 &  -1.38 &  0.153 &   -   &   -   &   -   &   -   &       BOES \\
             HD79452 &           G6III &   6.14 &   3.86 &  0.984 &  0.029 &   0.51 &  0.152 &  5041 &  2.40 & -0.69 &  1.57 &       BOES \\
       HD81040$^{a}$ &             G0V &   7.86 &   6.16 &  0.476 &  0.029 &   5.24 &  0.074 &  5795 &  4.71 & -0.04 &  0.80 &       BOES \\
             HD81146 &           K2III &   4.62 &   1.69 &  0.956 &  0.029 &   0.66 &  0.028 &  4449 &  2.20 & -0.04 &  1.51 &       BOES \\
       HD81688$^{a}$ &        K0III-IV &   5.56 &   3.06 &  0.812 &  0.029 &   0.87 &  0.073 &  4801 &  2.43 & -0.29 &  1.47 &       BOES \\
             HD82210 &        G4III-IV &   4.69 &   2.51 &  0.719 &  0.029 &   2.16 &  0.013 &  5335 &  3.44 & -0.19 &  1.26 &       BOES \\
       HD82885$^{a}$ &          G8IIIv &   5.54 &   3.69 &  0.563 &  0.029 &   5.25 &  0.008 &  5515 &  4.47 &  0.31 &  1.06 &       BOES \\
             HD83240 &           K1III &   5.17 &   2.66 &  0.836 &  0.016 &   0.77 &  0.106 &  4780 &  2.52 &  0.03 &  1.39 &       UVES \\
             HD84335 &           M3III &   5.14 &   0.34 &  1.126 &  0.015 &  -1.12 &  0.166 &   -   &   -   &   -   &   -   &       BOES \\
             HD89758 &           M0III &   3.15 &  -1.01 &  1.078 &  0.029 &  -1.11 &  0.083 &   -   &   -   &   -   &   -   &       BOES \\
             HD94600 &           K1III &   5.19 &   2.33 &  0.939 &  0.029 &   0.87 &  0.044 &   -   &   -   &   -   &   -   &       BOES \\
             HD95578 &           M0III &   4.83 &   0.81 &  1.229 &  0.029 &  -1.46 &  0.093 &   -   &   -   &   -   &   -   &       BOES \\
             HD99322 &           K0III &   5.38 &   3.04 &  0.812 &  0.011 &   0.74 &  0.058 &  4985 &  2.88 &  0.11 &  1.34 &       UVES \\
       HD99492$^{a}$ &             K2V &   7.70 &   5.26 &  0.452 &  0.029 &   6.41 &  0.059 &  4894 &  4.43 &  0.23 &  0.76 &       BOES \\
             HD99648 &          G8Iab: &   5.12 &   2.83 &  0.985 &  0.006 &  -1.12 &  0.124 &   -   &   -   &   -   &   -   &       UVES \\
            HD100029 &           M0III &   3.92 &  -0.11 &  1.018 &  0.029 &  -1.14 &  0.033 &   -   &   -   &   -   &   -   &       BOES \\
            HD100623 &             K0V &   6.10 &   4.02 &  0.404 &  0.001 &   6.19 &  0.008 &  5182 &  4.54 & -0.38 &  0.63 &       UVES \\
            HD101153 &           M4III &   5.22 &  -0.21 &  0.988 &  0.029 &  -0.76 &  0.119 &   -   &   -   &   -   &   -   &       BOES \\
      HD101501$^{a}$ &             G8V &   5.45 &   3.59 &  0.498 &  0.029 &   5.53 &  0.006 &  5507 &  4.66 & -0.02 &  0.84 &       BOES \\
      HD102195$^{a}$ &             K0V &   8.21 &   6.15 &  0.499 &  0.029 &   5.84 &  0.054 &  5252 &  4.39 &  0.06 &  0.76 &       BOES \\
            HD102212 &           M1III &   4.14 &   0.16 &  0.988 &  0.011 &  -0.64 &  0.035 &   -   &   -   &   -   &   -   &       UVES \\
            HD102224 &        K0.5IIIb &   3.86 &   0.99 &  0.979 &  0.029 &   0.10 &  0.020 &  4366 &  1.50 & -0.52 &  1.75 &       BOES \\
            HD102328 &           K3III &   5.43 &   2.63 &  0.922 &  0.029 &   1.32 &  0.043 &  4589 &  2.52 &  0.14 &  1.81 &       BOES \\
            HD104304 &            G8IV &   5.69 &   4.03 &  0.565 &  0.003 &   5.15 &  0.009 &  5572 &  4.37 &  0.27 &  0.97 &       UVES \\
            HD107325 &        K2III-IV &   5.69 &   3.10 &  0.767 &  0.029 &   2.21 &  0.038 &  4742 &  3.10 &  0.16 &  1.29 &       BOES \\
            HD107446 &         K3.5III &   3.73 &   0.31 &  1.046 &  0.002 &  -0.52 &  0.026 &  4144 &  1.26 & -0.26 &  1.74 &       UVES \\
            HD108225 &           G8III &   5.18 &   2.86 &  0.895 &  0.070 &   0.65 &  0.044 &   -   &   -   &   -   &   -   &       BOES \\
            HD108381 &           K1III &   4.52 &   1.81 &  0.898 &  0.029 &   0.96 &  0.021 &  4678 &  2.56 &  0.14 &  1.64 &       BOES \\
            HD110458 &           K0III &   4.83 &   2.27 &  0.820 &  0.021 &   0.99 &  0.028 &  4773 &  2.61 &  0.19 &  1.46 &       UVES \\
      HD110833$^{a}$ &             K3V &   7.16 &   4.78 &  0.471 &  0.029 &   6.29 &  0.021 &  4879 &  4.16 &  0.11 &  0.36 &       BOES \\
      HD111395$^{a}$ &             G5V &   6.43 &   4.64 &  0.558 &  0.029 &   5.28 &  0.017 &  5684 &  4.72 &  0.11 &  0.88 &       BOES \\
            HD112300 &           M3III &   3.44 &  -1.19 &  1.172 &  0.029 &  -0.49 &  0.029 &   -   &   -   &   -   &   -   &       BOES \\
            HD115383 &             G0V &   5.31 &   4.03 &  0.651 &  0.004 &   4.08 &  0.010 &  6176 &  4.37 &  0.21 &  1.26 &       UVES \\
      HD118203$^{a}$ &              K0 &   8.20 &   6.54 &  0.627 &  0.029 &   3.40 &  0.135 &  5650 &  3.70 &  0.16 &  1.01 &       BOES \\
            HD119149 &         M1.5III &   5.10 &   0.70 &  1.170 &  0.004 &  -0.89 &  0.092 &   -   &   -   &   -   &   -   &       UVES \\
            HD120052 &           M2III &   5.49 &   0.73 &  1.001 &  0.001 &  -1.20 &  0.176 &   -   &   -   &   -   &   -   &       UVES \\
            HD120933 &           K5III &   4.78 &  -0.01 &  1.019 &  0.001 &  -1.57 &  0.080 &   -   &   -   &   -   &   -   &       BOES \\
            HD121416 &           K1III &   5.98 &   3.44 &  0.882 &  0.022 &   1.01 &  0.087 &   -   &   -   &   -   &   -   &       UVES \\
            HD123657 &        M4.5:III &   5.10 &  -0.23 &  1.109 &  0.029 &  -0.90 &  0.089 &   -   &   -   &   -   &   -   &       BOES \\
            HD123934 &           M1III &   5.01 &   0.60 &  1.030 &  0.005 &  -0.76 &  0.098 &   -   &   -   &   -   &   -   &       UVES \\
      HD130215$^{a}$ &             K2V &   8.11 &   5.98 &  0.482 &  0.029 &   6.10 &  0.051 &  5298 &  4.83 &  0.18 &  0.90 &       BOES \\
      HD130307$^{a}$ &             G8V &   7.91 &   5.61 &  0.429 &  0.029 &   6.46 &  0.033 &  5124 &  4.89 & -0.12 &  0.57 &       BOES \\
            HD130328 &           M3III &   5.72 &  -0.54 &  1.175 &  0.005 &  -0.33 &  0.118 &   -   &   -   &   -   &   -   &       UVES \\
      HD130948$^{a}$ &             G1V &   5.99 &   4.46 &  0.579 &  0.029 &   4.68 &  0.013 &  6055 &  4.56 &  0.05 &  1.07 &       BOES \\
            HD131156 &             G8V &   4.68 &   1.97 &  0.461 &  0.029 &   5.54 &  0.007 &   -   &   -   &   -   &   -   &       BOES \\
      HD131511$^{a}$ &             K2V &   6.14 &   4.32 &  0.495 &  0.029 &   5.82 &  0.012 &  5331 &  4.70 &  0.12 &  0.81 &       BOES \\
            HD131873 &           K4III &   2.20 &   1.29 &  1.074 &  0.029 &  -0.83 &  0.010 &   -   &   -   &   -   &   -   &       BOES \\
            HD132813 &           M5III &   4.58 &  -0.96 &  1.163 &  0.029 &  -1.22 &  0.114 &   -   &   -   &   -   &   -   &       BOES \\
      HD135599$^{a}$ &             K0V &   7.06 &   4.96 &  0.443 &  0.029 &   6.05 &  0.024 &  5274 &  4.81 & -0.03 &  0.87 &       BOES \\
      HD137759$^{a}$ &           K2III &   3.46 &   0.67 &  0.855 &  0.029 &   0.99 &  0.007 &  4584 &  2.73 &  0.13 &  1.40 &       BOES \\
            HD138716 &            K1IV &   4.77 &   2.23 &  0.679 &  0.005 &   2.45 &  0.014 &  4804 &  2.94 &  0.01 &  1.24 &       UVES \\
            HD140573 &          K2IIIb &   2.80 &   0.15 &  0.887 &  0.006 &   1.01 &  0.009 &  4737 &  2.72 &  0.22 &  1.74 &       UVES \\
            HD140901 &            G7IV &   6.15 &   4.32 &  0.542 &  0.004 &   5.21 &  0.013 &  5652 &  4.53 &  0.15 &  0.80 &       UVES \\
            HD145206 &           K4III &   5.51 &   1.98 &  0.989 &  0.005 &  -0.38 &  0.128 &  4200 &  1.28 & -0.16 &  1.66 &       UVES \\
      HD145675$^{a}$ &             K0V &   6.76 &   4.71 &  0.518 &  0.029 &   5.53 &  0.013 &  5270 &  4.34 &  0.42 &  0.67 &       BOES \\
            HD147379 &             M1V &   8.66 &   4.95 &  0.369 &  0.029 &   8.51 &  0.022 &   -   &   -   &   -   &   -   &       BOES \\
            HD148291 &      K0II-IIICN &   5.36 &   2.41 &  1.135 &  0.032 &  -1.66 &  0.178 &  4545 &  1.64 & -0.09 &  2.05 &       UVES \\
            HD148387 &          G8IIIb &   2.87 &   0.58 &  0.803 &  0.029 &   0.61 &  0.006 &  4998 &  2.59 & -0.01 &  1.38 &       BOES \\
            HD148451 &           G5III &   6.74 &   4.43 &  0.831 &  0.055 &   0.23 &  0.161 &  4928 &  2.42 & -0.64 &  1.54 &       UVES \\
            HD149447 &           K6III &   4.28 &   0.39 &  1.161 &  0.058 &  -0.84 &  0.037 &   -   &   -   &   -   &   -   &       UVES \\
            HD149661 &             K2V &   5.91 &   4.04 &  0.465 &  0.002 &   5.95 &  0.008 &  5269 &  4.44 & -0.01 &  1.00 &       UVES \\
            HD156026 &             K5V &   6.44 &   3.47 &  0.358 &  0.000 &   7.55 &  0.008 &   -   &   -   &   -   &   -   &       UVES \\
            HD156274 &             G8V &   5.61 &   3.42 &  0.373 &  0.004 &   5.88 &  0.013 &   -   &   -   &   -   &   -   &       UVES \\
            HD156283 &            K3II &   3.31 &  -0.02 &  1.197 &  0.029 &  -2.01 &  0.030 &  4334 &  1.74 &  0.06 &  2.08 &       BOES \\
      HD156668$^{a}$ &              K2 &   8.57 &   6.00 &  0.439 &  0.029 &   6.62 &  0.046 &  4801 &  4.60 & -0.02 &  0.44 &       BOES \\
      HD160269$^{a}$ &            G0Va &   5.35 &   3.74 &  0.577 &  0.029 &   4.58 &  0.011 &  6004 &  4.55 &  0.05 &  0.99 &       BOES \\
      HD160346$^{a}$ &             K3V &   6.66 &   4.10 &  0.470 &  0.029 &   6.44 &  0.016 &  4873 &  4.26 & -0.02 &  0.69 &       BOES \\
            HD164058 &           K5III &   2.36 &  -1.16 &  1.101 &  0.029 &  -1.02 &  0.010 &  3928 &  1.54 &  0.12 &  1.84 &       BOES \\
      HD164922$^{a}$ &             K0V &   7.15 &   5.11 &  0.521 &  0.029 &   5.42 &  0.026 &  5297 &  4.47 &  0.16 &  0.60 &       BOES \\
      HD166620$^{a}$ &             K2V &   6.52 &   4.23 &  0.403 &  0.029 &   6.30 &  0.007 &  4974 &  4.59 & -0.19 &  0.05 &       BOES \\
      HD167042$^{a}$ &           K1III &   6.14 &   3.55 &  0.645 &  0.029 &   2.63 &  0.028 &  4908 &  3.26 &  0.03 &  1.07 &       BOES \\
            HD167818 &           K5III &   4.78 &   0.82 &  1.340 &  0.006 &  -2.17 &  0.167 &   -   &   -   &   -   &   -   &       UVES \\
            HD169916 &            K0IV &   2.98 &   0.33 &  0.778 &  0.013 &   1.07 &  0.008 &  4711 &  2.53 & -0.10 &  1.34 &       UVES \\
      HD183255$^{a}$ &             K3V &   8.15 &   5.57 &  0.412 &  0.029 &   6.15 &  0.040 &  4947 &  4.68 & -0.57 &  0.62 &       BOES \\
            HD188650 &        G1Ib-IIe &   5.93 &   4.11 &  1.429 &  0.029 &  -2.02 &  0.192 &   -   &   -   &   -   &   -   &       BOES \\
            HD189124 &           M6III &   4.76 &  -1.53 &  1.153 &  0.000 &  -1.08 &  0.083 &   -   &   -   &   -   &   -   &       UVES \\
            HD189763 &           M4III &   4.45 &  -0.83 &  1.072 &  0.005 &  -1.25 &  0.054 &   -   &   -   &   -   &   -   &       UVES \\
      HD190406$^{a}$ &             G0V &   5.92 &   4.39 &  0.565 &  0.029 &   4.66 &  0.014 &  6031 &  4.62 &  0.11 &  0.88 &       BOES \\
      HD190771$^{a}$ &            G5IV &   6.32 &   4.62 &  0.574 &  0.029 &   4.94 &  0.015 &  5949 &  4.72 &  0.20 &  1.15 &       BOES \\
            HD196171 &        K0III-IV &   3.27 &   0.70 &  0.815 &  0.013 &   0.86 &  0.012 &  4829 &  2.47 & -0.12 &  1.45 &       UVES \\
      HD198149$^{a}$ &            K0IV &   3.57 &   1.39 &  0.643 &  0.029 &   2.79 &  0.003 &  4928 &  3.30 & -0.16 &  0.88 &       BOES \\
            HD198357 &            K3II &   5.67 &   2.27 &  1.075 &  0.008 &  -0.67 &  0.129 &  4041 &  1.06 & -0.12 &  1.55 &       UVES \\
      HD199665$^{a}$ &          G6III: &   5.67 &   3.37 &  0.730 &  0.029 &   1.28 &  0.051 &  5047 &  3.10 &  0.07 &  1.18 &       BOES \\
            HD199951 &           G6III &   4.81 &   2.54 &  0.892 &  0.013 &   0.57 &  0.040 &  5145 &  2.80 & -0.05 &  1.46 &       UVES \\
            HD202320 &        K0II/III &   5.33 &   2.52 &  1.033 &  0.051 &  -1.16 &  0.107 &  4607 &  1.77 & -0.15 &  1.94 &       UVES \\
            HD206778 &            K2Ib &   2.55 &  -0.86 &  1.640 &  0.022 &  -4.10 &  0.078 &  4221 &  0.58 & -0.23 &  2.89 &       UVES \\
            HD209100 &             K5V &   4.83 &   2.24 &  0.391 &  0.001 &   7.03 &  0.002 &   -   &   -   &   -   &   -   &       UVES \\
            HD210066 &           M1III &   5.12 &   1.62 &  1.154 &  0.006 &  -0.47 &  0.062 &   -   &   -   &   -   &   -   &       UVES \\
            HD213080 &        M4.5IIIa &   4.14 &  -0.91 &  1.086 &  0.000 &  -0.90 &  0.051 &   -   &   -   &   -   &   -   &       UVES \\
            HD214665 &           M4III &   5.12 &  -0.16 &  1.064 &  0.029 &  -0.60 &  0.072 &   -   &   -   &   -   &   -   &       BOES \\
            HD214952 &           M5III &   2.07 &  -3.32 &  1.186 &  0.014 &  -1.61 &  0.052 &   -   &   -   &   -   &   -   &       UVES \\
            HD219215 &           M2III &   4.32 &  -0.10 &  1.094 &  0.011 &   0.30 &  0.120 &   -   &   -   &   -   &   -   &       UVES \\
      HD222404$^{a}$ &            K1IV &   3.38 &   1.04 &  0.679 &  0.029 &   2.62 &  0.012 &  4883 &  3.36 &  0.17 &  1.11 &       BOES \\
            HD224935 &           M3III &   4.41 &  -0.40 &  1.134 &  0.007 &  -1.31 &  0.170 &   -   &   -   &   -   &   -   &       UVES \\

\enddata
\tablenotetext{a}{ These BOES data have been published by \citet{kang11}.}
\tablenotetext{b}{ Spectral types are from the SIMBAD database. }
\tablenotetext{c}{ V magnitudes are taken from the $\it{Hipparcos}$ Catalogue \citep{leeuwen07}. }
\tablenotetext{d}{ K magnitudes are taken from the 2MASS All-Sky Catalogue \citep{cutri03}. }
\end{deluxetable}
\end{center}


\begin{deluxetable}{ccccrcc}
\footnotesize
\tabletypesize{\scriptsize}
\tablecaption{M67 \label{tbl_M67}}
\tablewidth{0pt}
\tablehead{
  \colhead{Sanders ID$^{\rm a}$} & \colhead{Width [\AA{}]$^{\rm b}$} & \colhead{Width [\AA{}]} & 
  \colhead{$m_{V}^{\rm c}$} & \colhead{$M_{V}^{\rm d}$}  &  \colhead{$(m-M)_{V}^{\rm e}$}  &  \colhead{$(m-M)_{V}$} \\
  \colhead{} & \colhead{} & \colhead{\citet{pace03}} & \colhead{} & \colhead{} & \colhead{} & \colhead{\citet{pace03} } 
} 
\startdata
  S978    & 1.089  & 1.080  &  9.72  & -0.936  & 10.656  & 10.926   \nl
  S1016  & 0.804  & 0.700  & 10.30  &  1.446  &  8.854  &  8.116   \nl
  S1221  & 0.894  & 0.854  & 10.76  &  0.613  & 10.147  & 10.131   \nl
  S1250  & 0.953  & 0.997  &  9.69  &  0.111  &  9.579  & 10.271   \nl
  S1479  & 0.908  & 0.868  & 10.55  &  0.491  & 10.059  & 10.048   \nl
\enddata
\tablenotetext{a}{ Sanders ID numbers were taken from \citet{sanders77}.}
\tablenotetext{b}{ Our measurement of the line width.}
\tablenotetext{c}{ The apparent magnitude from the $Hipparcos$ Catalogue \citep{leeuwen07}. }
\tablenotetext{d}{ The absolute magnitude calculated by our WBR.}
\tablenotetext{e}{ The distance modulus calculated with c and d.}
\end{deluxetable}

\begin{deluxetable}{ccc}
\footnotesize
\tabletypesize{\scriptsize}
\tablecaption{Correction for the Lutz-Kelker effect \label{tbl_LKE}}
\tablewidth{0pt}
\tablehead{
  \colhead{{$\frac{{\sigma}_{\pi}}{\pi}$}$^{\rm a}$} & \colhead{{$\Delta$M}$^{\rm b}$} & \colhead{Number of Stars}
} 

\startdata
0.0                &   0.0   & 0  \nl
0.0   $\sim$ 0.025 &  -0.01  & 81 \nl
0.025 $\sim$ 0.050 &  -0.02  & 23 \nl
0.050 $\sim$ 0.075 &  -0.06  & 13 \nl
0.075 $\sim$ 0.100 &  -0.11  & 8  \nl
\enddata
\tablenotetext{a}{ The ratio of parallax error to parallax.}
\tablenotetext{b}{ The correction for the absolute magnitude from \citet{lutz73}.}
\end{deluxetable}

\begin{deluxetable}{lcccccc}
\tabletypesize{\scriptsize}
\tablecaption{Application to M type stars\label{tbl_Mstar}}
\tablewidth{0pt}
\tablehead{
\colhead{HD name} & \colhead{Sp.type} & \colhead{\logw} & \colhead{\teff(V-K)} & \colhead{\logg(fit)} & \colhead{\teff(ref.)} & \colhead{\logg(ref.)} \\
\colhead{} & \colhead{} & \colhead{[km s$^{-1}$]} & \colhead{[K]} & \colhead{[dex]} & \colhead{[K]} & \colhead{[dex]}
}

\startdata

HD 89758   & M0 III & 1.92 & 3851.62 & 1.07 & 3700$^{\rm a}$ & 1.35$^{\rm a}$  \\
HD 101153  & M4 III & 1.88 & 3464.98 & 0.83 & 3452$^{\rm b}$ & 0.80$^{\rm b}$  \\
HD 102212  & M1 III & 1.88 & 3918.59 & 1.36 & 3738$^{\rm c}$ & 1.55$^{\rm c}$   \\
HD 123657  & M4.5 III & 1.93 & 3484.12 & 0.56 & 3261$^{\rm c}$ & 0.59$^{\rm c}$   \\

\enddata

\tablenotetext{a}{\citet{mallik98}}
\tablenotetext{b}{\citet{smith86}}
\tablenotetext{c}{\citet{koleva12}}

\end{deluxetable}

\end{document}